\documentstyle[11pt,mrs2003,epsfig]{article}
\begin{document} 
\title{Branes at Low Energy}

\author{Ph. Brax${}^1$, C. van de Bruck${}^2$, A.--C. Davis${}^{3}$ 
and C.S. Rhodes$^3$}
\affil{$^1$Service de Physique Th\'eorique, CEA-Saclay\\ 
F-91191, Gif/Yvette cedex, France \\
$^2$ Astrophysics Department, Oxford University, Keble Road \\
Oxford OX1 3RH, U. K.
\\
$^3$Department of Applied Mathematics and Theoretical Physics, 
Center for Mathematical Sciences,\\
University of Cambridge, Wilberforce Road, Cambridge CB3 0WA, U.K.}

\begin{abstract} 
Brane models at low energy are described by an effective action
involving gravity and two moduli fields associated with the brane positions.
The scalar-tensor nature of the effective theory and 
the  coupling to matter are explicitly spelt out. We discuss the post-Newtonian gravitational 
constraints on the models, some applications to quintessence and the variations of constants. 
\end{abstract} 
 
\section{Introduction} 
Extra--dimensional models open up new possibilities both in particle physics
and cosmology. In recent years, models inspired from M--theory or D--branes
have flourished revealing interesting features such as modifications
to general relativity  at high energy. We will focus on five dimensional
brane--world models  
(for a review see \cite{us}).

Brane--world models in five dimensions have an exact  low energy description
below the brane tension involving a four dimensional
effective action. Above the brane tension, non-conventional 
features such as the appearance of the famous $\rho^2$ term, i.e. the 
squared matter density, in the Friedmann equation precludes any attempt
to use a four dimensional approach. Below the brane tension, it has been
shown that the solutions of the five dimensional equations with boundary terms are identical to the solutions of the four dimensional equations derived from an effective action. This effective action is of the scalar--tensor type
with a universal coupling to matter on each brane. 
In section two we will present the effective action and the coupling to matter. Post--Newtonian constraints from solar system experiments will be used
 to restrict the coupling of the bulk scalar field to the brane 
and the radion value now. In section three we will mention how exponential
quintessence can be obtained by detuning the brane tension. In Section four,
some comments on the variations of constants in brane models will be presented.   
\section{The Effective Action} 
We consider brane--world models in 5d with a bulk scalar field $\psi$\cite{br1,br2}. 
The bulk action consists of two terms which describe 
gravity and the bulk scalar field dynamics:
\begin{equation}
S_{\rm bulk} = \frac{1}{2\kappa_5^2} \int d^5 x \sqrt{-g_5}
\left( {\cal R} - \frac{3}{4}\left( (\partial \psi)^2 + U \right)\right).
\end{equation}
Further, our setup contains two branes. One of these branes has a
positive tension, the other brane has a negative tension.  They are
described by the action
\begin{eqnarray}
S_{\rm brane 1} &=& -\frac{3}{2\kappa_5^2}\int d^5x \sqrt{-g_5} U_B
\delta(z_1), \label{b1} \\
S_{\rm brane 2} &=& +\frac{3}{2\kappa_5^2}\int d^5x \sqrt{-g_5} U_B \delta(z_2) \label{b2}.
\end{eqnarray}
In these expressions, $z_1$ and $z_2$ are the (arbitrary) positions of
the two branes, $U_B$ is the superpotential; $U$, the bulk potential
energy of the scalar field, is given by
\begin{equation}
U = \left(\frac{\partial U_B}{\partial \psi}\right)^2 - U_B^2.
\end{equation}
We will also include the Gibbons--Hawking boundary term for each
brane, which have the form
\begin{equation}
S_{\rm GH} = \frac{1}{\kappa_5^2}\int d^4 x \sqrt{-g_4} K,
\end{equation}
where $K$ is the extrinsic curvature of the individual branes.
We impose a $Z_2$--symmetry at the position of each brane. 

The solution of the system above can be derived from 
BPS--like equations of the form
\begin{equation}
\frac{a'}{a}=-\frac{U_B}{4},\ \psi'=\frac{\partial U_B}{\partial \psi},
\end{equation}
where $'=d/dz$ for a metric of the form
\begin{equation}\label{background}
ds^2 = dz^2 + a^2(z)\eta_{\mu\nu}dx^\mu dx^\nu.
\end{equation}
We will particularly focus on the case where the superpotential is an 
exponential function:
\begin{equation}\label{potential}
U_B=4k e^{\alpha \psi}.
\end{equation}
The values  $\alpha =1/\sqrt 3,-1/\sqrt {12}$ were obtained in a theory
with supergravity in singular spaces. The solutions read
\begin{equation}\label{scale}
a(z)=(1-4k\alpha^2z)^{\frac{1}{4\alpha^2}},
\end{equation}
while the scalar field solution is
\begin{equation}\label{psi}
\psi = -\frac{1}{\alpha}\ln\left(1-4k\alpha^2z\right).
\end{equation}
In the $\alpha\to 0$ we retrieve the AdS profile
\begin{equation}
a(z)=e^{-kz}.
\end{equation}
Notice that in that case the scalar field decouples altogether. Also, 
notice that there is a singular point in the bulk at 
$z_* = 1/4k\alpha^2$, for which the scale factor vanishes. 

In the following we will discuss the moduli space approximation.
At low energy below the brane tension this leads to an exact description of the brane system.  Two of the moduli of the
system are the brane positions, i.e. in the solution above the
brane positions are arbitrary.  In the moduli space approximation,
these moduli are assumed to be space--time dependent.  We denote the
position of brane 1 by  $z_1 = \phi(x^\nu)$ and the position of
brane 2 by $z_2 = \sigma(x^\mu)$. We consider the case where the
evolution of the brane is slow. This means that in constructing the
effective four--dimensional theory we neglect terms like $(\partial
\phi)^3$.

In addition to the brane positions, we need to include the graviton 
zero mode, which can be done by replacing $\eta_{\mu\nu}$ with a 
space--time dependent tensor $g_{\mu\nu}(x^\mu)$. Thus, we have 
two scalar degrees of freedom, namely the positions of the two 
branes which we will denote by  $\phi(x^\mu)$ and $\sigma(x^\mu)$, 
and the graviton zero mode $g_{\mu\nu}$.  

The effective action is obtained by substituting the metric ansatz in the 
5d action, allowing fluctuations of the brane locations. The result to
second order in a derivative expansion reads\cite{br2} 
\begin{eqnarray}
S_{\rm MSA} = \int d^4 x \sqrt{-g_4}\left[ f(\phi,\sigma) {\cal R}^{(4)} 
+ \frac{3}{4}a^2(\phi)\frac{U_B(\phi)}{\kappa_5^2}(\partial \phi)^2 
- \frac{3}{4} a^2(\sigma)\frac{U_B}{\kappa_5^2}(\sigma)(\partial \sigma)^2 \right].
\end{eqnarray}
where the effective gravitational constant is 
\begin{equation}
f(\phi,\sigma) = \frac{1}{\kappa_5^2} \int^{\sigma}_{\phi} dz a^2 (z).
\end{equation}
It is convenient to go to redefine the moduli fields
\begin{eqnarray}
\tilde \phi^2 &=& \left(1 - 4k\alpha^2 \phi\right)^{2\beta}, \label{posia1}\\
\tilde \sigma^2 &=& \left(1-4k\alpha^2 \sigma\right)^{2\beta} \label{posia2},
\end{eqnarray}
with 
$
\beta = \frac{2\alpha^2 + 1}{4\alpha^2};
$
and
\begin{eqnarray}
\tilde \phi &=& Q \cosh R, \label{posib1} \\
\tilde \sigma &=& Q \sinh R \label{posib2}.
\end{eqnarray}
This diagonalises the kinetic terms of the moduli.

The gravitational coupling  can be made constant in the Einstein frame where
\begin{equation}
\tilde g_{\mu\nu} = Q^2 g_{\mu\nu}.
\end{equation}
leading to the effective action
\begin{eqnarray}
S_{\rm EF} &=& \frac{1}{2k\kappa^2_5(2\alpha^2 + 1)} 
\int d^4x \sqrt{-g}\left[ {\cal R} -  \frac{12\alpha^2}{1+2\alpha^2}
\frac{(\partial Q)^2}{Q^2} - \frac{6}{2\alpha^2 + 1}(\partial R)^2\right].
\end{eqnarray}
where the gravitational constant is 
\begin{equation}
16\pi G = 2k\kappa_5^2 (1+2\alpha^2).
\end{equation}
Notice that the two moduli fields $S=\ln Q$ and $R$ are massless fields.
There are two special points in the moduli space. When $R$ vanishes, the
second brane hits the bulk singularity while $Q=0$ corresponds to the
collision of the two branes. 
We will see that these special points play a particular role in the dynamics
of the brane system. 

Let us now discuss the coupling to matter. Matter couples to the induced
metric on the branes. As the bulk is warped the coupling to matter on the first and the second brane differ drastically. In the Einstein frame the matter
action reads
\begin{equation}
S_m^{(1)} = S_m^{(1)}(\Psi_1,A^2(Q,R)g_{\mu\nu}) \hspace{0.5cm} {\rm and}
\hspace{0.5cm} S_m^{(2)} = S_m^{(2)}(\Psi_2,B^2(Q,R)g_{\mu\nu}),
\end{equation}
where $\psi_{1,2}$ are the the matter fields on the first and the second brane respectively. The coupling constants $A$ and $B$ are given by 
\begin{equation}
A=Q^{-\frac{\alpha^2\lambda}{2}}(\cosh R)^{\frac{\lambda}{4}},\ B=Q^{-\frac{\alpha^2\lambda}{2}}(\sinh R)^{\frac{\lambda}{4}}
\end{equation}
where $\lambda=4/(1+2\alpha^2)$
Notice that when converging to the singularity, the coupling of the second
brane $B$ vanishes.

In the following we will concentrate on matter on the first brane.
As matter couples to both gravity and the moduli, one expects
strong deviations from general relativity. In particular, solar system experiments would detect the presence of massless moduli unless the Eddington parameter $\gamma -1\approx -2 \theta$ is close enough to one where
\begin{equation}
\theta= \frac{4}{3}\frac{\alpha^2}{1+2\alpha^2}+ \frac{\tanh^2 R}{6(1+2\alpha^2)}.
\end{equation}
More precisely one must impose $\theta \le 10^{-4}$\cite{will}
implying that
\begin{equation}
\alpha\le 10^{-2},\ R\le 0.2
\end{equation}
Now it happens that the evolution of the moduli $R$ in the matter dominated
era follows
\begin{equation}
R = R_1 \left(\frac{t}{t_0}\right)^{-1/3} + 
R_2 \left(\frac{t}{t_0}\right)^{-2/3}  
\end{equation}
implying that  general relativity is a late time attractor, i.e. the brane
fly far apart from each other.
Hence, solar system experiments are satisfied now for a large range of initial
conditions. Of course a large variation of the moduli field since nucleosynthesis is not acceptable either. This leads to another bound
\begin{equation}
R_{BBN}\le 0.4
\end{equation}
Different time evolutions for the field $R$ can be seen in figure 1. 

\section{Moduli Quintessence}

So far we have considered the moduli to be strictly massless. We will
see that one can generate a potential for such moduli, leading to exponential quintessence.
Let us first consider that a potential term is present on the first brane. This leads
to the effective term in the action
\begin{equation}
\int d^4 x \sqrt{-g_4} \left[ a^4(\phi)V\right]
\end{equation}
with $a^4(\phi) = \tilde \phi^{4/(1+2\alpha^2)}$.
Going to the Einstein frame results in a new potential 
\begin{equation}
V_{\rm eff}(Q,R) = Q^{-8\alpha^2/(1+2\alpha^2)}(\cosh
R)^{4/(1+2\alpha^2)} V.
\end{equation}
Notice the extra contributions from the moduli fields.
%
%
\begin{figure}  
\vspace*{1.25cm}  
\begin{center}
\epsfig{figure=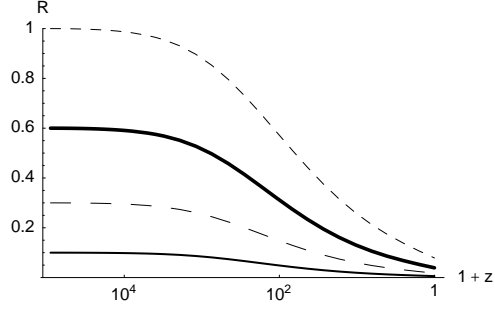,width=6.5cm}  
\end{center}
\vspace*{0.25cm}  
\caption{The field $R$ rolls down towards zero and general relativity  is retrieved now  
} 
\end{figure}

The existence of the moduli is tightly related to the tuning between the potential $U_B$ and
the bulk potential $U$. When one detunes the brane tension $U_B\to T U_B$, the moduli
pick up a potential
\begin{equation}
V = \frac {6(T-1)k}{\kappa_5^2} e^{\alpha \psi}.
\end{equation}
 Expressed in terms of $S$ and $R$ we have
\begin{equation}\label{poti}
V_{eff}(S,R) = \frac{6(T-1)k}{\kappa_5^2} e^{-12\alpha^2 S/(1+2\alpha^2)}
\left(\cosh R \right)^{(4-4\alpha^2)/(1+2\alpha^2)}.
\end{equation}
in the Einstein frame. 
As expected the $R$ dependence of the potential for small $\alpha$ implies that $R$ is attracted
towards zero.
For $R$ small the effective potential becomes
\begin{equation}\label{poti2}
V_{eff}(S) = \frac{6(T-1)k}{\kappa_5^2} e^{-12\alpha^2 S/(1+2\alpha^2)}.
\end{equation}
corresponding to a quintessence potential.
Notice that here we have obtained a coupled quintessence model \cite{amendola}
where both the baryons and cold dark matter couple to the moduli fields.

The exponential potential admits an attractor\cite{liddle} with a scale factor
\begin{equation}
a=a_0 t^{\frac{(1+2\alpha^2)}{3\alpha^2}}
\end{equation}
which is accelerating when $\alpha <1$. On the attractor, the scalar field
dominates with $\Omega_S =1$.
The equation of state on the attractor
is given by
\begin{equation}
w=-1+\frac{4\alpha^2}{(1+2\alpha^2)}
\end{equation}
This result has already been obtained using the 5d equations of motion\cite{br1}.
Notice that for the Randall-Sundrum case $\alpha=0$, one gets de Sitter branes.
Another interesting case is $\alpha=-1/\sqrt {12}$ where $w=-5/7$ (supergravity case).
The five dimensional picture  corresponds to a brane moving  at constant speed
in the bulk written in terms of conformal coordinates.
Using the constraints from solar system experiments we find that $w \le -0.96$, i.e. extremely close to the de Sitter $(w=-1)$ case.

Of course when coupled to matter the previous attractor can only be reached if one fine--tunes the value of $S_{now}$ such that 
\begin{equation}
\Omega_{S_{now}}\approx 0.7
\end{equation}
i.e. not on the attractor yet.

\section{Conclusions} 
Brane-- world models at low energy are described by an effective action involving gravity and two moduli fields. Such a scalar--tensor theory would lead
to large deviations from general relativity if the coupling of the bulk scalar field to the branes and one of the moduli were not small now. 
We have exhibited a quintessence potential for the moduli obtained by detuning the brane tension on the positive tension brane.  
Another consequence of the cosmological evolution of moduli is the
time--dependence of fine--structure  constant provided one couples the electromagnetic kinetic terms 
to the bulk scalar field\cite{palma}
\begin{equation}
\frac{\delta\alpha_{EM}}{\alpha_{EM}}=-\alpha_{EM}\frac{16\alpha\beta}{9(1+2\alpha^2)}\ln (1+z)
\end{equation}
 The coupling $\beta$ has to be of the same order of magnitude
as the brane coupling to the bulk scalar field in order to both comply with solar system experiments and results on the variation of the fine--structure constant\cite{webb}.
Finally the presence of moduli fields coupled to cold dark matter and baryons
leads to significant modifications to the CMB spectrum\cite{rhodes}.

\vfill 
\end{document}